# Exponential versus linear amplitude decay in damped oscillators

**M. I. Molina**
*Departamento de Física, Facultad de Ciencias, Universidad de Chile, Casilla 653, Santiago, Chile*
*mmolina@uchile.cl*

**Abstract:** We comment of the widespread belief among some undergraduate students that the amplitude of any harmonic oscillator in the presence of any type of friction, decays exponentially in time. To dispel that notion, we compare the amplitude decay for a harmonic oscillator in the presence of (i) viscous friction and (ii) dry friction. It is shown that, in the first case, the amplitude decays exponentially with time while in the second case, it decays *linearly* with time.

Key words: oscillator, viscous friction, dry friction

**PACS:** 01.55, 46.30M, 46.30P

## 1. Introduction

The problem of oscillatory motion is, without any doubt, one of the main topics in physics, from elementary up to advanced courses. An understanding of this motion is also relevant in many areas outside physics, including chemistry, biology, engineering, medical research and economics, to name a few. In physics, students encounter oscillatory behavior in classical mechanics, electricity, optics and later, in quantum mechanics.

When considering adding some `realistic' features to a simple harmonic oscillator, such as damping effects, it is a common practice in many textbooks, to consider only the case of an oscillator in the presence of a viscous force, where the resistive force is $-\gamma \mathbf{v}$, where $\gamma$ is the viscous coefficient and $\mathbf{v}$ is the velocity of the oscillator. As shown in many textbooks[1], this leads to an exponential decrease of the oscillator's amplitude with time. One important case which is not normally treated in standard textbooks is the oscillator in the presence of sliding (dry) friction, even though this type of friction is present in all aspects of our life. Because of this, some students are led to believe that, in the absence of driving forces, the motion of a harmonic oscillator in the presence of any friction, will decay exponentially in time.

Here we try to dispell that notion by comparing the amplitude decay for a harmonic oscillator in the presence of viscous friction and dry friction.

## 2. Viscous friction

A simple example of this is an object like a pendulum, or a block attached to a spring, that oscillates in the presence of a drag force provided, for instance, by the surrounding air. At low speeds, the resistive force has the form $-\gamma \mathbf{v}$, where $\gamma$ is the viscous coefficient and is $\mathbf{v}$ the velocity of the oscillator. In one dimension, Newton's equation reads:

$$m \frac{d^2 x(t)}{dt^2} = -k\, x(t) - \gamma \frac{dx(t)}{dt}. \qquad (1)$$

We look for a solution of the form $\exp(-\lambda t)$. After replacing in Eq.(1), we obtain an equation for $\lambda$:

$$\lambda^2 + b\lambda + \omega^2 = 0 \qquad (2)$$

where $b = \gamma/m$ and $\omega = \sqrt{k/m}$. Thus, we obtain two solutions for $\lambda = -(b/2) \pm i\sqrt{\omega^2 - (b/2)^2}$, and the oscillator's displacement has the general form $x(t) = \exp(-(b/2)t)[A\cos(\Omega t) + B\sin(\Omega t)]$, where $\Omega = \sqrt{\omega^2 - (b/2)^2}$. To keep things simple, let us take the initial conditions $x(0) = x_0$ and $(dx/dt)_0 = 0$. Also, we will assume that $b \ll \omega$, that is, we are in the ``underdamped regime'', where the system performs many oscillations before coming to a rest. Thus, we obtain

$$x(t) = \exp[-(b/2)t](\cos(\Omega t) + \sin(\Omega t)), \quad (3)$$

and the oscillator amplitude is then exponentially damped with time. In particular, this means that the maximum amplitude points $x_n$ where the velocity is zero, will also decay exponentially in time or cycle number n. Let us show explicitly how $x_n$ depends on n: From (3), the oscillator's velocity $v(t) = dx(t)/dt$ is found to be

$$v(t) = \frac{\omega^2 x_0}{\Omega} \exp[-(b/2)t] \sin(\Omega t). \quad (4)$$

Thus, $v = 0$ happens at $t_n = n\pi/\Omega$, and the amplitude $A_n$ will be

$$A_n = |x(t_n)| = \exp[-(b/2)n\pi/\Omega]. \quad (5)$$

From (5) it is clear that the envelope of the oscillatory motion decays *exponentially* with the half-cycle number n. This is readily apparent in the envelope of the oscillatory motion depicted in Fig.1.

## 3. Sliding (dry) friction

This motion has been analyzed by several authors with varying degrees of sophistication[2,3,4,5]. We will restrict ourselves here to a very elementary treatment, which can be followed by a student with little or no calculus. The most common example of this motion is that of a block resting on a rough table and attached to a horizontal spring. Let us assume equal values for the static and kinetic coefficients of friction. Newton's equation in this case reads:

$$m \frac{d^2 x(t)}{dt^2} = -k x(t) - \text{sgn}\left[\frac{dx}{dt}\right] \mu m g, \quad (6)$$

where $\mu$ is the coefficient of dry friction, and sgn(x) is defined to be +1 for x > 0, -1 for x < 0 and zero for x=0. Thus, the frictional force is constant in magnitude but acts in the opposite direction to the velocity. Suppose we start from rest with an initial amplitude $A_0$. For the block to move, the force provided by the spring $kA_0$ must be greater that the magnitude of the static friction, $\mu m g$. This means, $A_0 > \mu m g \equiv A_c$. Let us assume that $A_0 \gg A_c$, so that the system will perform several oscillations before coming to rest. After releasing the block with initial amplitude $A_0$, it will traverse a distance $A_0 + A_1$, where it will come to rest momentarily for the first time. Conservation of energy requires that the initial potential energy $(1/2) k A_0^2$ be equal to the new potential energy $(1/2) k A_1^2$ plus the energy lost to friction $\mu m g (A_0 + A_1)$:

$$(1/2) k A_0^2 = (1/2) k A_1^2 + \mu m g (A_0 + A_1). \quad (7)$$

After rewriting this as

$$(1/2) k (A_0^2 - A_1^2) = \mu m g (A_0 + A_1) \quad (8)$$

and, after simplifying, one obtains:

$$A_1 = A_0 - 2(\mu m g / k) = A_0 - 2 A_c. \quad (9)$$

The block will swing back, provided $k A_1 > \mu m g$. If this is so, at the end of the cycle, we will have:

$$A_2 = A_1 - 2 A_c = A_0 - 4 A_c. \quad (10)$$

This continues for a while, and after n of such half-cycles,

$$A_n = A_0 - 2 n A_c. \quad (11)$$

Eventually however, the oscillator's maximum amplitude will decrease below $A_c$, and the spring will not be strong enough to overcome the dry friction and subsequently, all motion will stop.

As Eq.(11) clearly shows, the decay of the maximum amplitude of the oscillator is *linear* with the half-cycle number (or time), and is vividly illustrated in Fig.1. This stands in marked contrast with the previous case of the oscillator subjected to viscous friction.

Now, as to the question of why most Physics textbooks do not consider dry friction and focus on viscous friction only when dealing with a damped harmonic oscillator, it probably has to do with the rather awkward form of the dry friction term: $-\text{sgn}(dx/dt)\mu\, m\, g$, which is not amenable to a simple closed-form solution, as in the viscous case.

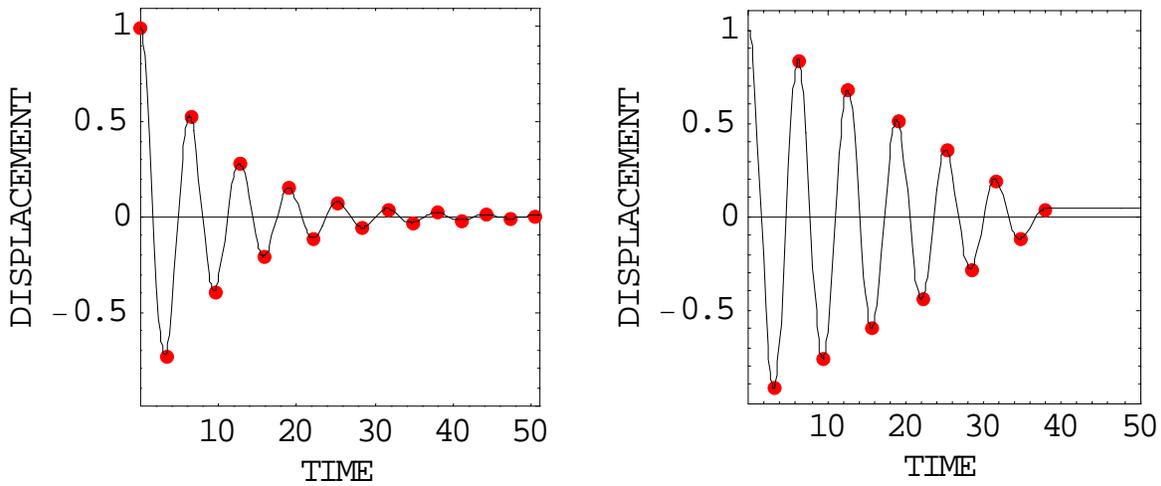

Fig. 1. Amplitude decay of a harmonic oscillator in the presence of viscous friction (left) and dry friction (right). ( Parameters used: m = 1, k = 1, g = 1, γ = 0.1 and μ = 0.04 ).